\documentclass[prl,aps,twocolumn,superscriptaddress]{revtex4}

\usepackage{bm}
\usepackage{textcomp}
\usepackage{dcolumn}
\usepackage{graphicx}
\usepackage{amsfonts}
\usepackage{amsmath}
\usepackage{amssymb}

\begin{document}

\title{
Impurity assisted nanoscale localization of plasmonic excitations in graphene}

\author{Rodrigo A. Muniz}
\email{rmuniz@usc.edu}
\affiliation{Department of Physics and Astronomy, University of Southern California, Los Angeles, CA 90089-0484}

\author{Hari P. Dahal}
\affiliation{Theoretical Division and Center for Integrated Nanotechnology, Los Alamos National Laboratory, Los Alamos, New Mexico 87545}

\author{A.V. Balatsky}
\email{avb@lanl.gov}
\affiliation{Theoretical Division and Center for Integrated Nanotechnology, Los Alamos National Laboratory, Los Alamos, New Mexico 87545}

\author{Stephan Haas}
\affiliation{Department of Physics and Astronomy, University of Southern California, Los Angeles, CA 90089-0484}

\date{\today}

\begin{abstract}
The plasmon modes of pristine and impurity doped graphene are calculated, using a real-space theory which determines the non-local dielectric
response within the random phase approximation.
A full diagonalization of the polarization operator is performed, allowing the extraction of all its poles.
It is demonstrated how  impurities induce the formation of localized modes which are absent in pristine graphene.
The dependence of the spatial modulations over few lattice sites and frequencies of the localized plasmons on the electronic filling and impurity strength is discussed.
Furthermore, it is shown that the chemical potential and impurity strength can be tuned to control target features of the localized modes.
These predictions can be tested by scanning tunneling microscopy experiments.
\end{abstract}

\maketitle

\textbf{Introduction}:
Graphene is an allotropic form of carbon, in which atoms are arranged in a two-dimensional honeycomb lattice.
Due to its thin crystal structure it is a promising material for a wide range of applications.\cite{geim2007}
The prospects for its use in nanotechnology are further reinforced by the success of multiple fabrication alternatives.\cite{novoselov2005,wilson2006}
Ab-initio calculations have shown that graphene is a gapless semiconductor, for which the valence and conduction bands meet at the Fermi energy.
At this point the energy dispersion of the quasi-particle excitations has been found to be linear.\cite{wallace1947}
Experimentally, the system can be tuned into a metallic regime by adjusting the chemical potential using a gate voltage.\cite{novoselovI2005}
Due to these unique properties there is considerable current interest in the electronic response of this
material.\cite{castroneto2009,gusynin2005,zhang2005,katsnelson2006,ohishi2007,herbut2008,williams2009}

While much theoretical research has focused on the ground state electronic properties of graphene, so far relatively few
attempts have been made to study the collective excitations of this system. Nevertheless, there are
some recent studies of the plasmonic modes in graphene using the linearized band structure\cite{wang2007,hwang2007}
and a full tight-binding band structure calculation.\cite{stauber2009}
The effects of a fully gapped band\cite{pyatkovskiy2009} and doping\cite{jablan2009} have also been explored.
Some of the above mentioned papers have worked out the dispersion relation of plasmons.

Recent STM experiments have shown that graphene is intrinsically disordered.\cite{brar2007}
In order to determine its technological usefulness, it is therefore important to understand the effects of disorder on its electronic properties.
Obviously, in the presence of impurities the system looses its translational symmetry, and
it is not known what types of localized modes form around them, e.g. dipole, quadrupole, radial etc.
Much recent research has focused on the effects of impurities on the ground state properties of graphene \cite{pereira2007, bena2008,wehling2007,
wehling2009,mallet2007, tan2007}, but it is equally important to understand how impurities affect collective modes.
In this letter, we examine the consequences of the presence of impurities on the plasmonic modes in graphene.
Our main focus is to obtain and control localized plasmons. The analysis of these features requires the determination of their spatial profile, which - to
the our best knowledge - has not yet been discussed in the literature on graphene.

\textbf{Model}:
The calculation of the dielectric response in metallic materials is conventionally performed using continuum-field Mie theory.\cite{mie}
However, such a semi-empirical continuum description breaks down beyond a certain degree of roughness, introduced by atomic length
scales.\cite{kresin}
Recently, a self-consistent quantum-mechanical approach was developed, which accounts for the non-locality of the dielectric response
function.\cite{Ilya} Using this technique, the identification of plasmons was accomplished by scanning the frequencies of the modes with strongest induced fields.
However, it does not provide full information of all plasmon excitations supported by the system.
In this contribution, we generalize this approach. The polarization operator is diagonalized, providing all its poles. Thus complete information of the
plasmon excitations is obtained, including the local spectral densities of states.

We model the electronic structure of graphene using a one-band tight-binding Hamiltonian,
\begin{equation}
H_{0} =-t\sum_{<a,b>}\left( c_{a}^{\dagger}c_{b} + c_{b}^{\dagger} c_{a}\right) + \sum_{a}U_{a}c_{a}^{\dagger}c_{a}  -  \mu \sum_{a} c_{a}^{\dagger}c_{a} \text {,}
\label{hamiltonian}
\end{equation}
where $t=2.7eV$ is the hopping parameter and $\mu$ is the chemical potential. $U_{0}$ is the magnitude of the impurity potential, ${\mathbf{x}_0}$ corresponds to its location, and $\sigma$ denotes its spatial spread.
$U_{a}$ is then the impurity potential felt at the other sites $\mathbf{x}_a \neq {\mathbf{x}_0}$, parametrized by $U_a=U_{0}\exp \left( \frac{-\left\vert {\mathbf{x}}_a -{\mathbf{x}_0} \right\vert ^{2}}{2\sigma ^{2}} \right)$. In this article, the range of the impurity potential $\sigma$ will be only few lattice sites.
$H_{0}$ is diagonalized numerically, providing the eigenstates $\left\vert \Psi_{\alpha}^{0} \right\rangle$ and eigenvalues $E_{\alpha }^{0}$.

The direct Coulomb interaction is considered as a perturbation,
\begin{equation}
H=H_{0}+V=H_{0}+\sum_{abmn}V_{abmn}c_{a}^{\dagger}c_{b}^{\dagger}c_{m}c_{n},
\end{equation}
with
\begin{equation}
V_{abmn}=\frac{e^{2}}{2}\int \!d\mathbf{x}\int \!d\mathbf{x}^{\prime }\;
\frac{\varphi _{a}^{\ast }(\mathbf{x})\varphi _{b}^{\ast }(\mathbf{x}
^{\prime })\varphi _{m}(\mathbf{x}^{\prime })\varphi _{n}(\mathbf{x})}{
\left\vert \mathbf{x}-\mathbf{x}^{\prime }\right\vert },
\end{equation}
where $\varphi_{j}(\mathbf{x})$ is the $p_z$ orbital at site  $j$.
In the lattice basis the induced charge is given by
\begin{equation}
\delta {\rho }\left( \mathbf{x}\right)
=\sum_{ab}\varphi _{a}^{\ast}(\mathbf{x}) \delta {\rho}_{ab} \varphi_{b}(\mathbf{x}) \notag
=\sum_{b}\varphi_{b}^{\ast }(\mathbf{x})\delta {\rho}_{bb} \varphi_{b}(\mathbf{x}) \text{.}
\label{eqn:overlap}
\end{equation}
Here the overlap of $p_z$ orbitals at different sites is neglected $\varphi_{a}^{\ast}(\mathbf{x}) \varphi_{b}(\mathbf{x}) = {\delta}_{ab}
\varphi_{b}^{\ast}(\mathbf{x}) \varphi_{b}(\mathbf{x})$.
The linear response charge equation in this basis is
\begin{equation}
\delta {\rho }_{ab}(\omega )=\sum_{mn}\Pi _{ab,mn}\left( \omega \right) {\phi }_{mn}^{Ext}(\omega )\text{.}
\end{equation}
Within the random phase approximation (RPA), the polarization operator is then obtained via
$ \Pi (\omega ) = \Pi^{0}\left( \omega \right) \left( 1-V \Pi^{0}\left(
\omega \right) \right)^{-1} $
where  $\Pi ^{0}\left( \omega \right)$ is the polarization operator of the non-interacting system. In the basis of eigenstates $\left\vert \Psi_{\alpha}^{0}
\right\rangle$, $\Pi ^{0}\left( \omega \right)$ can be written as
\begin{equation}
\Pi_{\alpha \beta ,\gamma \delta }^{0}\left( \omega \right) = \delta_{\alpha \gamma} \delta_{\beta \delta} \frac{n_{\alpha}^{0} - n_{\beta}^{0}} {E_{\alpha }^{0}-E_{\beta }^{0} - \omega} \text{.}
\end{equation}
This fourth rank tensor can be regarded as a matrix acting on the vectors ${\phi }_{ab}$, and thus can be represented as the product
\begin{equation}
\Pi^{0}\left( \omega \right) = \Delta n \left( \Delta E-\omega I\right) ^{-1} \text{,}
\end{equation}
where $\Delta n$ and $\Delta E$ are diagonal matrices
\begin{eqnarray}
\Delta n_{\alpha \beta ,\gamma \delta } &=&\delta _{\alpha \gamma }\delta
_{\beta \delta }\left( n_{\alpha }^{0}-n_{\beta }^{0}\right)   \notag \\
\Delta E_{\alpha \beta ,\gamma \delta } &=&\delta _{\alpha \gamma }\delta
_{\beta \delta }\left( E_{\alpha }^{0}-E_{\beta }^{0}\right) \text{.}
\end{eqnarray}
The polarization of the interacting system can then be expressed as
\begin{equation}
\Pi \left( \omega \right) = \Delta n\left( \Delta E - \omega I - V\Delta n\right)^{-1} \text{.}
\end{equation}
Hence the plasmons correspond to charge densities $\delta \rho = \Delta n{\phi}$ such that
\begin{equation}
\left( \Delta E-\omega I-V\Delta n\right) {\phi }=0 \text{.}
\label{condition}
\end{equation}
When the matrix $\Delta E-\Delta n V$ is diagonalized,
the polarization has  poles at $\omega = \lambda_{b}$ for each eigenvalue $\lambda_{b}$ of $\Delta E-\Delta n V$.

The retarded Green's function $\Pi ^{R}\left( \omega \right) = \Pi \left( \omega +i0^{+}\right)$ gives the spectral density function $A\left( \omega \right)
=-\frac{1}{\pi } Im \Pi^{R}\left( \omega \right)$, which is expressed as
\begin{equation}
A\left( \omega \right) = -\Delta n\delta \left( \Delta E-\omega I-V\Delta n\right) \text{.}
\end{equation}
Due to the delta function $A_{\alpha \beta ,\gamma \delta} \left( \omega \right)$ is non-zero only for plasmonic frequencies.
The only spatial profiles it can display are those of the plasmons $\Delta n_{ab,mn}{\phi }_{mn}$ with ${\phi}_{mn}$ satisfying $\left( \Delta E-\omega I-V\Delta n\right) {\phi }=0$.
In order to plot the plasmon density of states the representation $\delta \left( \omega - {\omega}_{\alpha} \right) \to \frac{\gamma}{\pi \left[ {\left( \omega
- {\omega}_{\alpha} \right)}^2 + {\gamma}^2 \right] }$ is used, where $\gamma$ is chosen to be $0.05 eV$.
Then the plasmon density of states $tr A(\omega)$ can be written as
\begin{equation}
tr A\left( \omega \right) = \sum_{\alpha }A_{\alpha }\left( \omega
\right)
=\sum_{\alpha } A_{\alpha }\frac{\gamma }{\pi \left[ {\left( \omega -{\omega }_{a}\right) }^{2}+{\gamma }^{2}\right]}
\end{equation}
$A_{\alpha}$ will be referred to as the strength of the mode
\begin{equation}
A_{\alpha} = ({\phi }^{\alpha })^{T} \Delta n {\phi}^{\alpha} = \sum_b({\phi }^{\alpha })_{bb}^{T} (\Delta n {\phi}^{\alpha})_{bb} \text{,}
\label{densit}
\end{equation}
where $\Delta n {\phi}^{\alpha}= \delta \rho$ is the charge profile of the mode $\alpha$.
Using Eq. \ref{densit}, the plasmon is considered to be localized when the major contribution to the sum comes from sites around the impurity.
The following discussion of our results is focussed on such localized modes.
The plasmon density of states $tr A\left( \omega \right)$ can be accessed experimentally through scanning tunneling experiments that reveal either direct or inelastic tunneling signatures associated with plasmons or related lifetime effects. \cite{brar2010}

For the numerical simulation a finite-size realization of the graphene lattice with 96 sites is considered. An impurity affects approximately one hexagon of the
honeycomb lattice, as shown in Fig. \ref{fig:dos}(a).
Periodic boundary conditions are used to eliminate the boundary modes.
LAPACK routines are used for the numerical diagonalization of  $H_0$ and $\Pi \left( \omega \right)$. \cite{Lapack}

\textbf{Results}:
\begin{figure}[h]
\begin{tabular}{r}
	\begin{tabular}{lr}
		{\includegraphics[width=3.8cm]{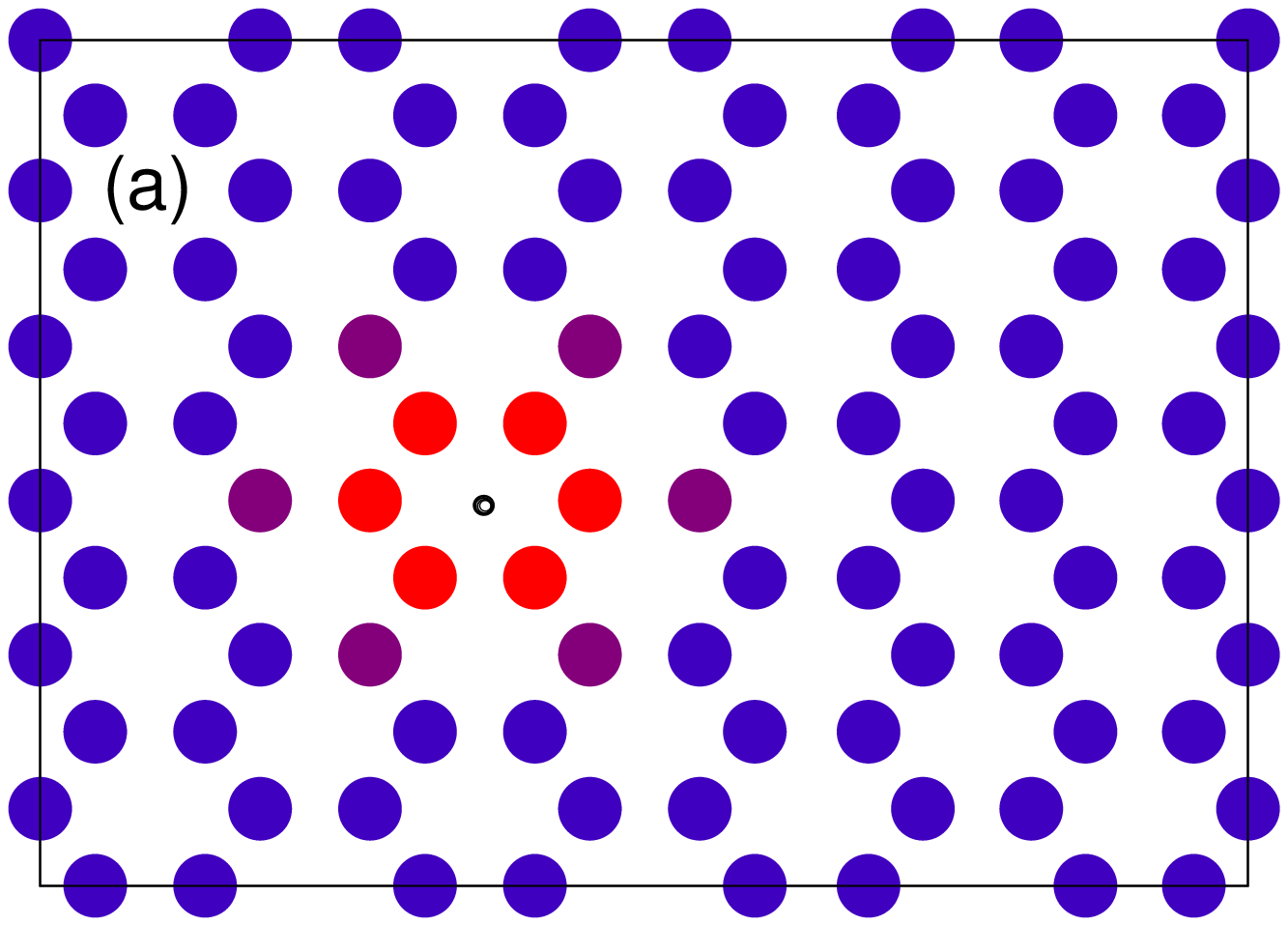}} &
		{\includegraphics[width=4.2cm]{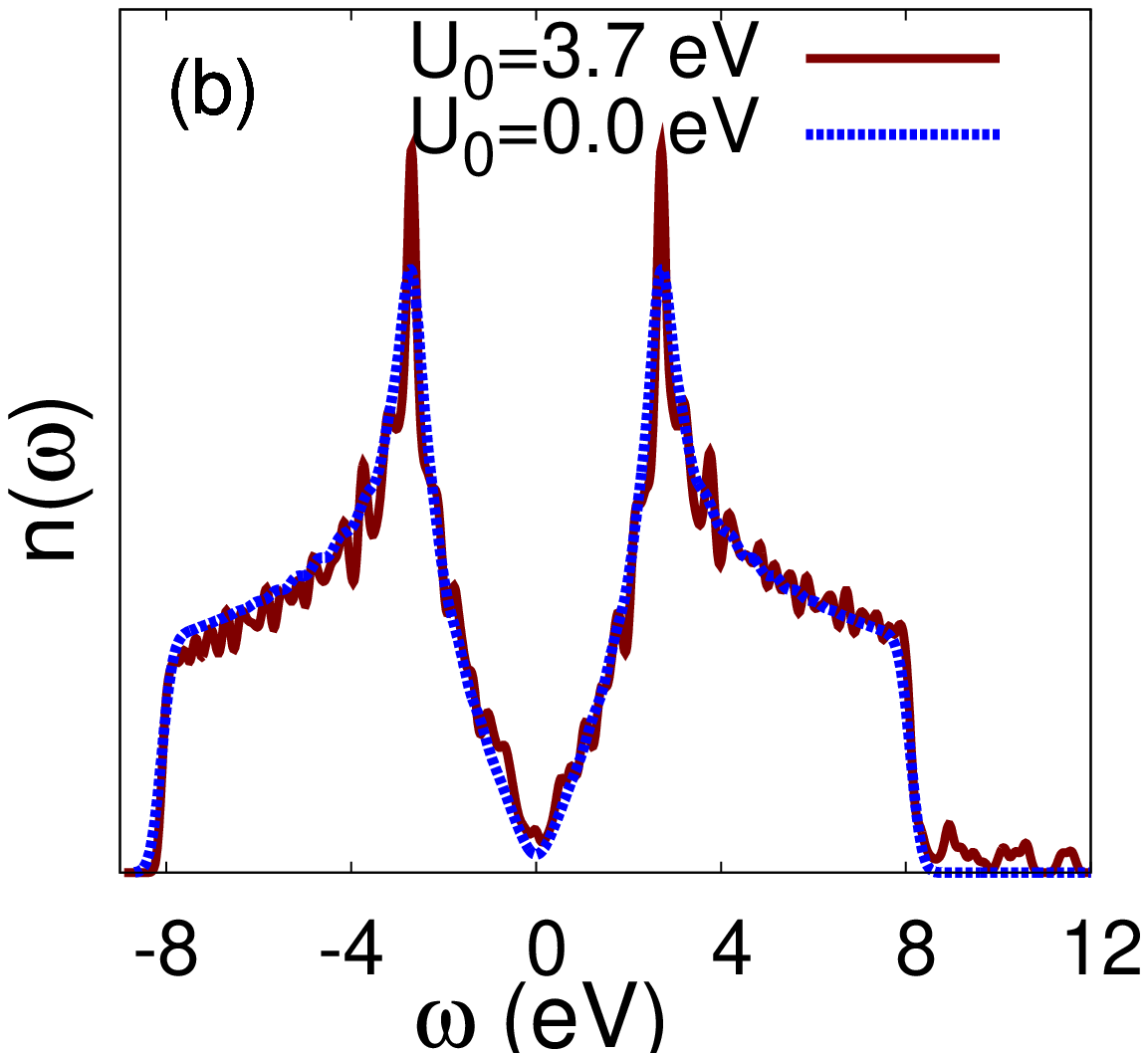}} \\
	\end{tabular} \\
	\\
	\\
	{\includegraphics[width=8.0cm]{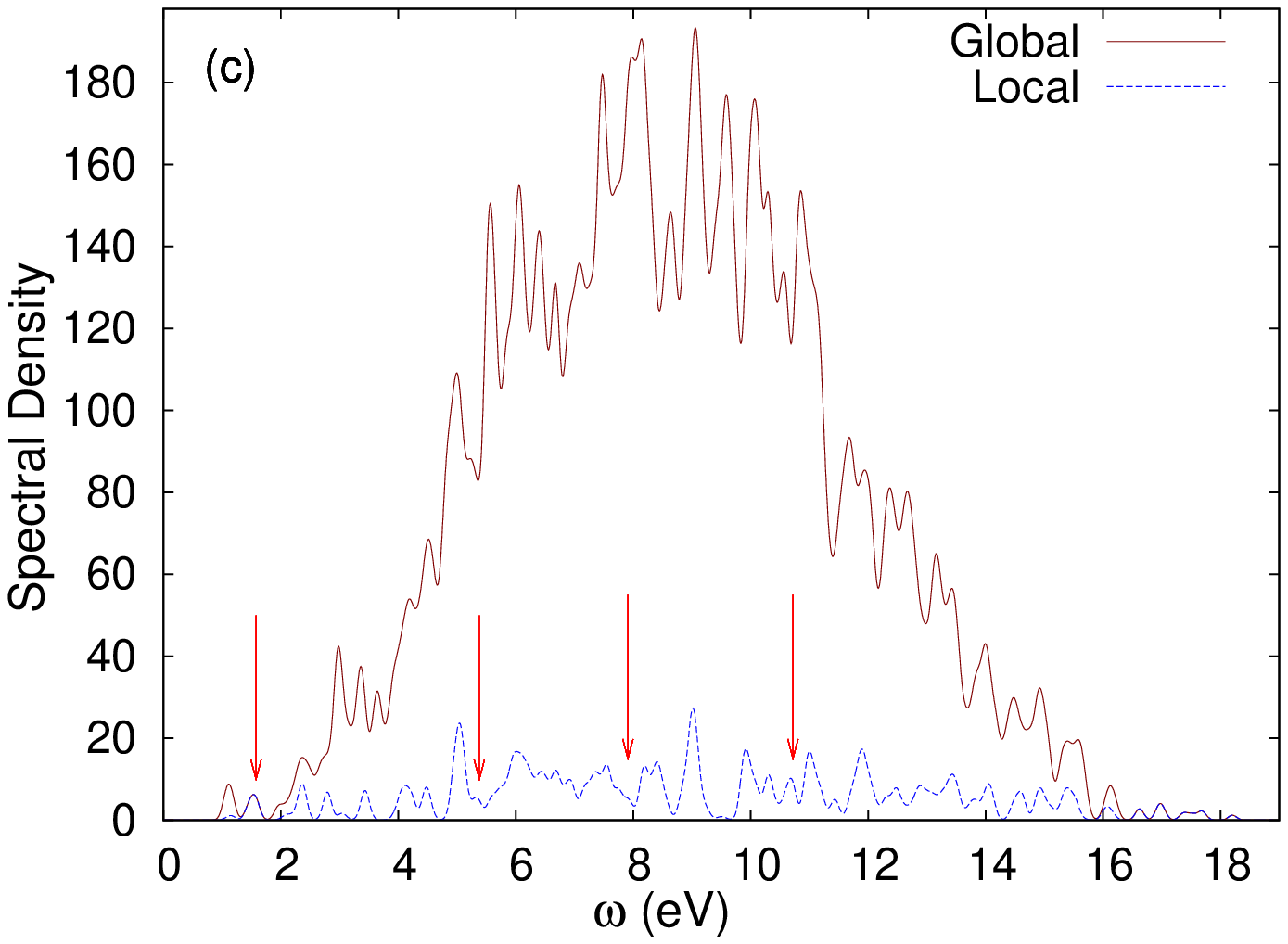}} \\
\end{tabular}
\vspace{1cm}
\caption{(Color online) (a) Single impurity  on the graphene lattice. Blue corresponds to $0 eV$ and red to $3.7 eV$.
The black dot indicates the center of the impurity $\mathbf{x}_0$.
(b) Single particle density of states $n(\omega) = -\frac{1}{\pi} Im ~ tr \mathcal{G}^{0} (\omega) = \sum_{\beta} \delta(\omega - E^0_{\beta})$ for pure($U_0
=0$) and doped($U_0 = 3.7 eV$) graphene, both with $\mu =0$.
(c) Spectral density of plasmons in graphene with $\mu =0$ and $U_0 =3.7 eV$.
``Global" corresponds to the density of states $tr A(\omega) = \sum_{all ~ \alpha} A_{\alpha}(\omega)$, and ``Local" is the spectral density of plasmons
localized around the impurity $tr A(\omega) = \sum_{local ~ \alpha} A_{\alpha}(\omega)$, here $\alpha$ runs through localized modes only.
The arrows indicate the modes shown in Fig.\ref{fig:modes}.}
\label{fig:dos}
\end{figure}

In Fig. \ref{fig:dos}(b) the single particle densities of states of pure and impurity doped graphene are shown. They feature two characteristic singularities
around $\pm 3.8 eV$ and a V-shaped dip at the Fermi energy. Please note that there is a tail of states in the doped system beyond the regular band width and also
some additional states around $\omega = 0$. In Fig.\ref{fig:dos}(c) the spectral density of all plasmon and the spectral density of localized plasmons are shown
for $\mu =0$ and $U_0 = 3.7 eV$. The global density of plasmons can be understood in terms of the single particle density of states shown in
Fig.\ref{fig:dos}(b). There are very few single particle states available around $\omega = 0$. Then there are two peaks around $\omega = \pm 3.8 eV$.
And, at the extrema of the band the single particle density of states becomes small again. This implies a quadratically increasing spectral density at small
$\omega$, a maximum around $\omega = 3.8-(-3.8) = 7.6 eV$, and a small impurity-dominated contribution around $\omega = 16 eV$ (see Fig. \ref{fig:dos}(c)).
Localized modes occur throughout the available frequency spectrum. Non-local plasmonic modes are most abundant around $8 eV$, because of the larger phase space
provided by the van-Hove singularities. The highest energy modes are all localized, i.e. the spectral density of local plasmons equals the total spectral density
above $16 eV$. This is a consequence of the fact that the high energy plasmons are excitations from the lowest energy states, which tend to be uniformly spread
throughout the lattice, to the tail of highest (localized) single-particle states, thus generating a very localized charge density profile.
The high energy values for these localized plasmonic modes are set by the very localized nature of the impurity problem.
Had the impurity potential range been much larger, on the scale of $10 nm$, one would expect those localized plasmonic modes with much smaller energy, on the scale of few eV.

\begin{figure}[h]
\begin{center}
	\begin{tabular}{lr}
		{\includegraphics[width=4.0cm]{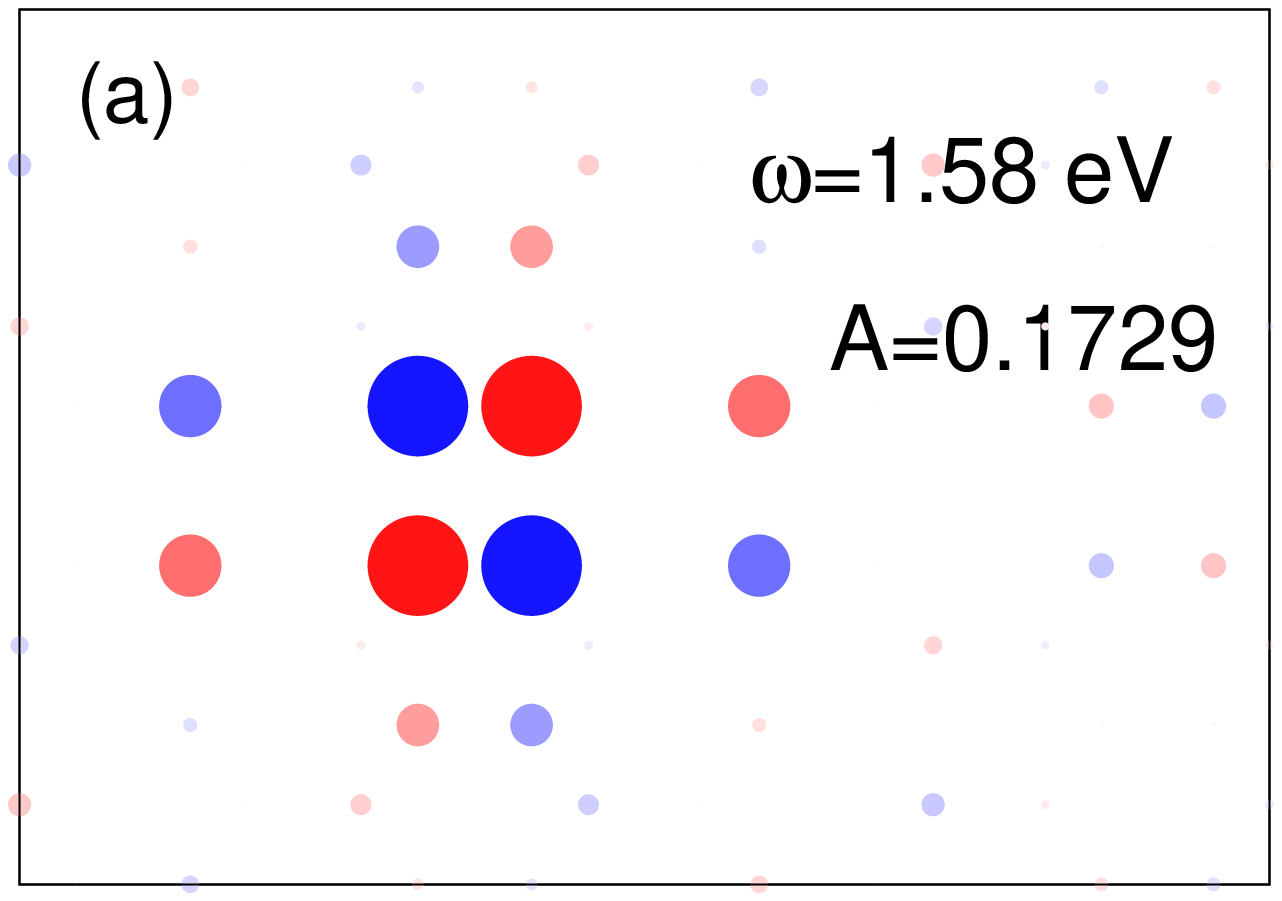}} &
		{\includegraphics[width=4.0cm]{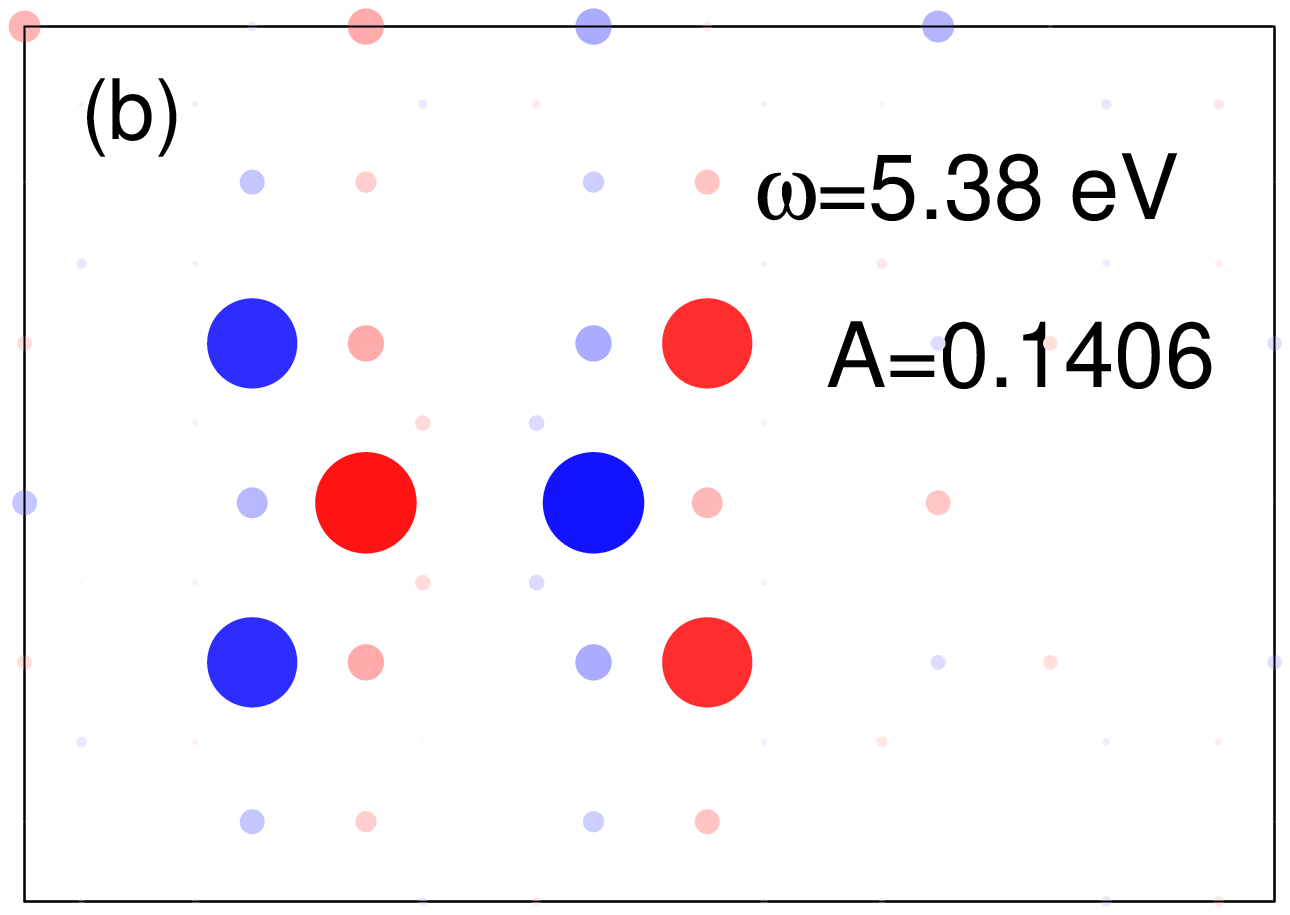}} \\
		{\includegraphics[width=4.0cm]{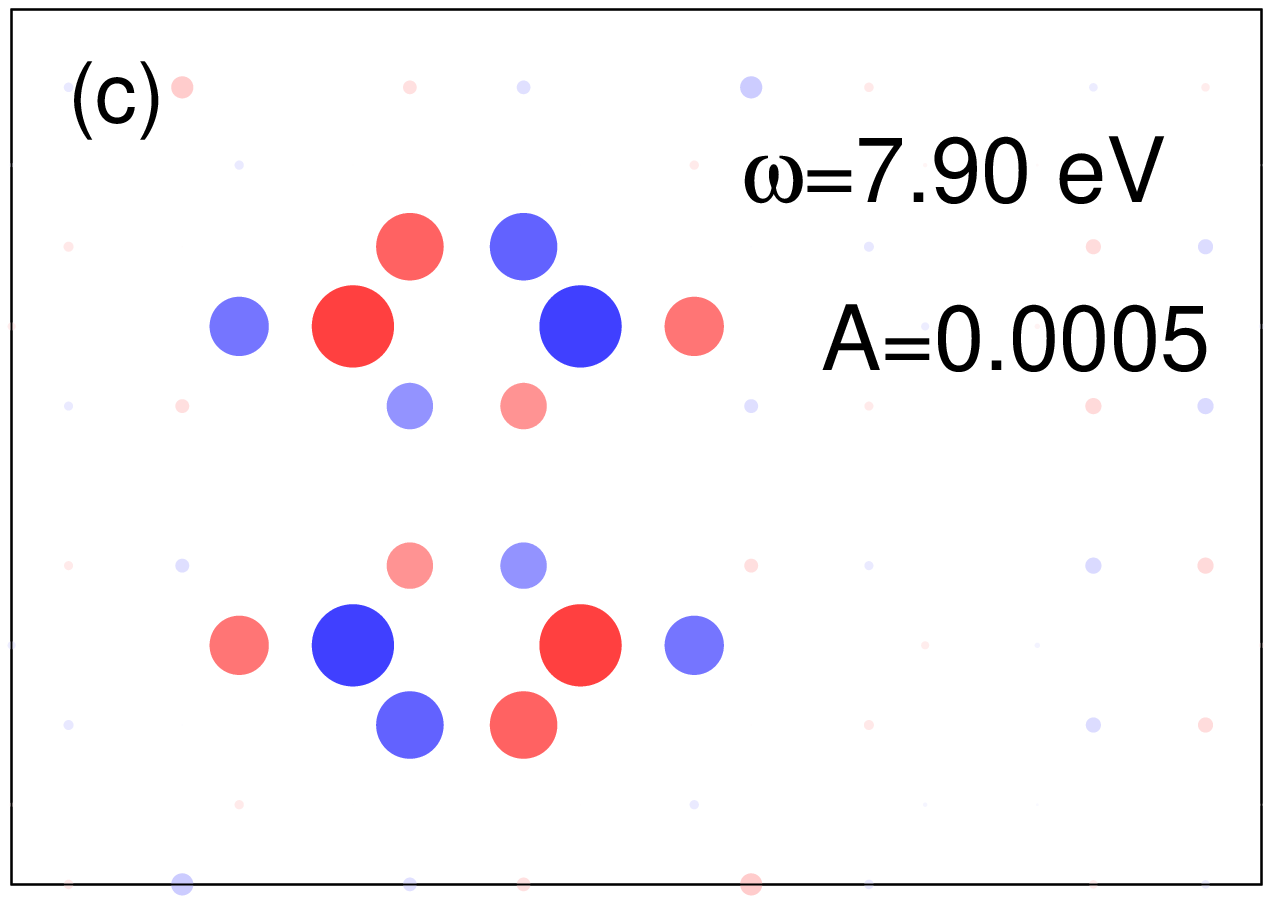}} &
		{\includegraphics[width=4.0cm]{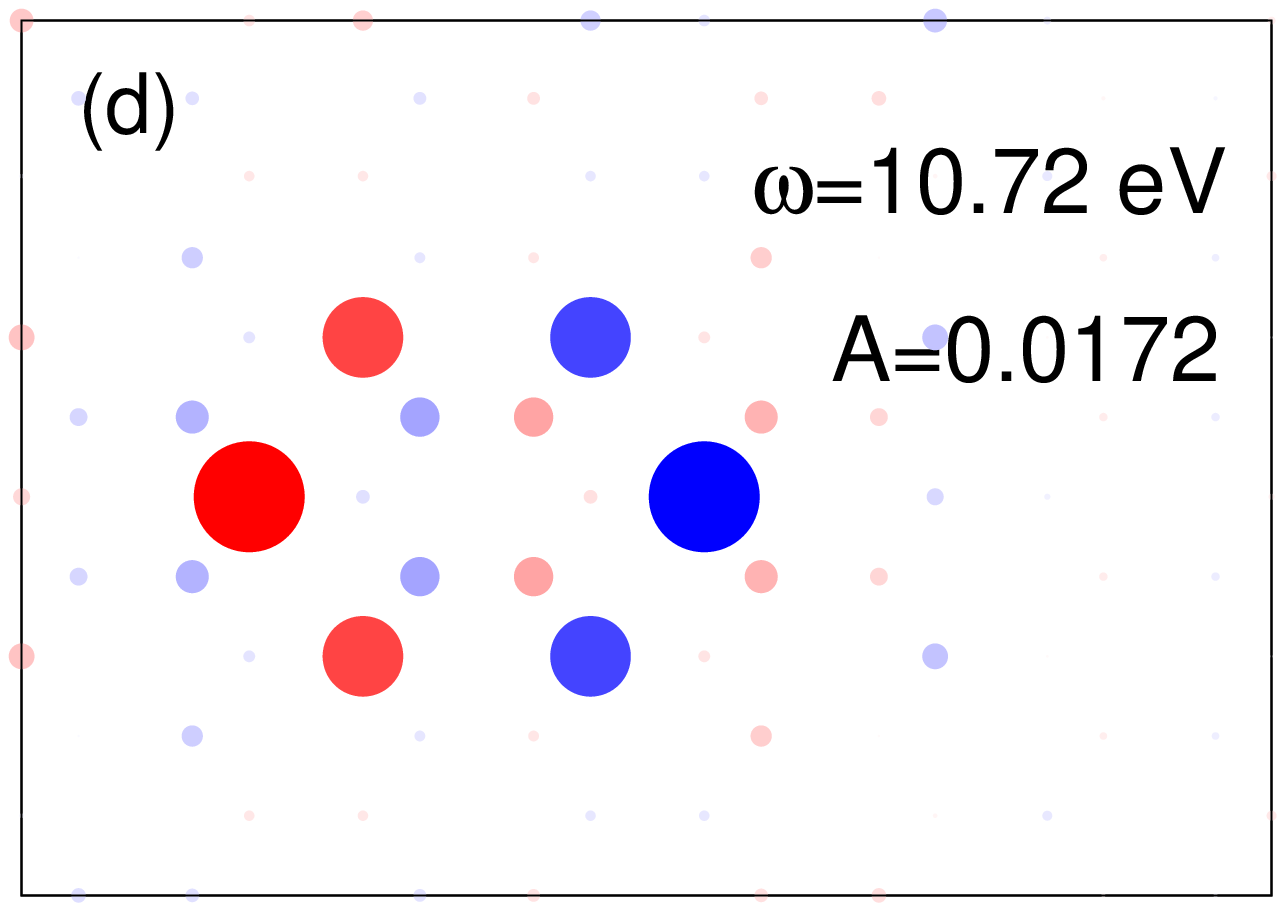}} \\
	\end{tabular}
\end{center}
\caption{(Color online) (a)-(d) Charge density profiles of some localized plasmons. Red (blue) corresponds to negative (positive) charges. The plots show the
induced charge $(\Delta n {\phi}^{\alpha})_{bb}$ for the mode $\alpha$ on site $b$. The frequencies $\omega$ and strengths $A_{\alpha}=({\phi}^{\alpha })^{T}
\Delta n {\phi}^{\alpha}$ of each mode are also indicated.}
\label{fig:modes}
\end{figure}

In Figs. \ref{fig:modes}(a)-(d) the spatial profiles of some selected localized modes are shown.
These are representatives for the diversity of localized modes. It is seen that some modes have strong dipole characteristics, whereas others have a strong
quadrupole component. This point highlights the importance of resolving the spatial distribution of the induced charge for all possible modes. Previous methods would
only be capable of detecting selected few modes with strong dipole moments.\cite{Ilya}

\begin{figure}[h]
\includegraphics[width=9cm]{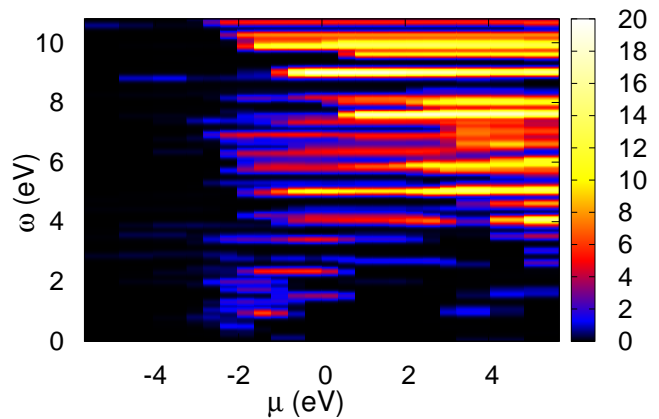}
\caption{(Color online) Dependence of the spectral density $tr A(\omega)$ of localized plasmon modes on the chemical potential $\mu$ and the frequency $\omega$.
The impurity potential is kept fixed at $U_0 =3.7 eV$. The spectral density is shown via the color scale.}
\label{fig:chempot}
\end{figure}

The dependence of the intensity of the localized plasmons (for a fixed impurity potential) on the chemical potential is shown in Fig. \ref{fig:chempot}.
For larger chemical potentials, more modes are present at lower energies. This is a consequence of the occupation of single particle states closer to those
localized modes at higher energies. When the chemical potential has opposite sign with respect to the impurity potential, there is some spectral density at low
energies. These modes stem from the extra single particle states around $\omega = 0$ in doped graphene, seen in Fig. \ref{fig:dos}(b).
This shows that the chemical potential, which can be experimentally controlled through a gate voltage, is an important tuning parameter for achieving certain target modes.
A limiting aspect of this property though is that the frequency of the modes does not change very significantly.
This happens because the chemical potential does not change the single particle states but only their occupation.
In this sense, the chemical potential does not allow one to control the frequency of localized plasmons but only their existence.
As it will be shown below the impurity potential $U_0$ changes the frequency of the localized modes more effectively.
We emphasize that Fig. \ref{fig:chempot} shows localized plasmonic excitations.\cite{footnote}

\begin{figure}[h]
\includegraphics[width=9cm]{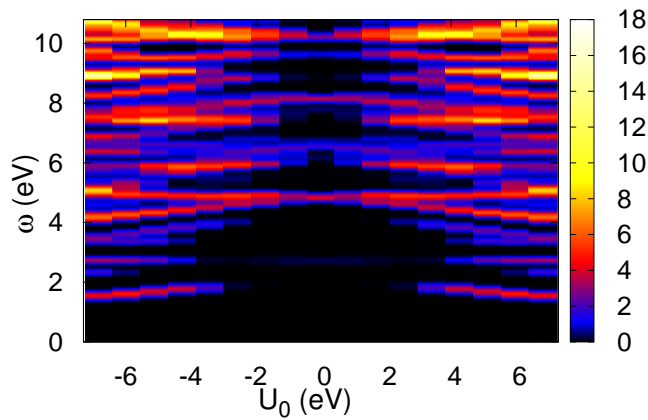}
\caption{(Color online) Dependence of the spectral density $tr A(\omega)$ for localized plasmons on the impurity potential magnitude $U_0$ and on the frequency
$\omega$. The chemical potential is kept fixed $\mu =0$. The spectral density is shown via the color scale.}
\label{fig:imppot}
\end{figure}

In order to change the frequency of modes, one can vary the impurity potential, because it actually affects the single particle states.
Fig. \ref{fig:imppot} shows that for an appropriate change of  $U_0$ the frequency of some localized plasmons can be tuned by about $1 eV$.
Note also that the impurity potential should be at least $2 eV$ in order to obtain a significant intensity of localized modes, and some modes are strong only
within a specific range of the impurity potential (see e.g. the features at $\omega \sim 6.0 eV$ or $\omega \sim 7.5 eV$).
The impurity potential considerably changes the spectral intensity of localized plasmons and can therefore be used as a tuning parameter to achieve targeted spectral properties, especially when combined with variations of the chemical potential.
This confirms should be expected since it is already known that the impurity changes the single particle density of states. \cite{pereira2007, bena2008, wehling2007}
Another striking feature displayed in this figure is the symmetry of the intensity of the modes with respect to the sign of the impurity potential.
This is a consequence of the single particle spectrum being symmetric relative to $\mu=0$. Since the plasmon involves electrons and holes, its properties do not
depend on the sign of the impurity potential.

\textbf{Conclusions}:
We have introduced an RPA approach which resolves the real space structure of plasmonic modes in graphene. This method was used to show that impurities induce the formation of nanoscale localized plasmonic excitations in graphene sheets. The spatial profile, i.e. dipole vs. multipole, of the modes was found to vary strongly with the particular resonance. Furthermore, their frequency and amplitude can be tuned by varying the strength of the impurity potential.  We also studied the effect of varying the chemical potential on these modes. It was found that the chemical potential can be used to turn them on and off, but it does not affect their frequency.
This theoretical study is a first step in exploring surface enhancement phenomena in graphene which may proof useful for nanoscale technologies such as molecular sensing.

\begin{acknowledgments}
\textbf{Acknowledgments}:
We thank Ming-Chak Ho, Noah Bray-Ali, Yung-Ching Liang and James Gubernatis for useful conversations. We also acknowledge financial support by the US-DOE through the BES and LDRD funds, and grant number DE-FG02-06ER46319. The numerical computations were carried out on the USC-HPC cluster.

\end{acknowledgments}


\begin{thebibliography}{9}
\bibitem{footnote} There is no contradiction with the well known fact that the usual bulk plasma frequency for metals depends on the number of electrons in the system.
In bulk plasmons almost every electron participate in the excitation while for a localized plasmon there is a definite amount of electrons involved. Adding an extra electron in the system directly affects a bulk plasmon mode while localized ones are only disturbed through the perturbative interaction.
\end{thebibliography}
\end{document}